\pdfoutput=1
\documentclass[]{spie}
\usepackage{epsfig}
\usepackage[]{graphicx}
\usepackage{subfig}
\usepackage{soul}
\usepackage{color}
\usepackage{url} 
\usepackage{booktabs}
\usepackage{siunitx}
\usepackage{float}
\usepackage{multirow}
\usepackage{color, colortbl}
\definecolor{LightCyan}{rgb}{0.88,1,1}
\definecolor{Gray}{rgb}{0.9,0.9,0.9}
\definecolor{Gray2}{rgb}{0.97,0.97,0.97}
\definecolor{Gray3}{rgb}{0.93,0.93,0.93}
\definecolor{White}{rgb}{1,1,1}
\usepackage{float}
\restylefloat{table}

\newcommand{\degree}{$^{\circ}\,$}

\title{Overcoming the effect of pupil distortion in multiconjugate adaptive optics}

\author[a]{Marcos A. van Dam}
\author[b]{Yolanda Mart\'in Hernando}
\author[b]{Miguel N\'u\~nez Cagigal}
\author[b]{Luzma M. Montoya}

\affil[a]{Flat Wavefronts, 21 Lascelles Street, Christchurch 8022, New Zealand}
\affil[b]{Instituto de Astrof\'isica de Canarias,  C/ V\'ia L\'actea s/n 38205, La Laguna, Spain}

\authorinfo{Further author information: send correspondence to Marcos van Dam: marcos@flatwavefronts.com}

\begin{document}
\maketitle

\begin{abstract}

Multiconjugate adaptive optics (MCAO) systems have the potential to deliver diffraction-limited images over much larger fields of view than traditional single conjugate adaptive optics systems. In MCAO, the high altitude deformable mirrors (DMs) cause a distortion of the pupil plane and lead to a dynamic misregistration between the DM actuators and the wavefront sensors (WFSs). The problem is much more acute for solar astronomy than for night-time observations  due to the higher spatial sampling of the WFSs and DMs, and the fact that the science observations are often made through stronger turbulence and at lower elevations. The dynamic misregistration limits the quality of the correction provided by solar MCAO systems. 

In this paper, we present PropAO, the first AO simulation tool (to our knowledge) to model the effect of pupil distortion. It takes advantage of the Python implementation of the optical propagation library PROPER. PropAO uses Fresnel propagation to propagate the amplitude and phase of an incoming wave through the atmosphere and the MCAO system. The resulting wavefront is analyzed by the WFSs and also used to evaluate the corrected image quality. We are able to reproduce the problem of pupil distortion and test novel non-linear reconstruction strategies that take the distortion into account. PropAO is shown to be an essential tool to study the behavior of the wavefront reconstruction and control for the European Solar Telescope. 

%
%
%

\end{abstract}
\keywords{multiconjugate adaptive optics, solar, wavefront reconstruction, pupil distortion}


\section{Introduction}
\label{sec:intro}

Multiconjugate adaptive optics (MCAO) uses two or more deformable mirrors (DMs) at different conjugate altitudes in order to correct the three-dimensional structure of turbulence and hence increase the corrected field-of-view relative to conventional single conjugate adaptive optics (SCAO). 

MCAO has enjoyed some success in night time astronomy, with the MAD demonstrator at the VLT\cite{MAD} and GeMS\cite{RigautGeMS,NeichelGeMS} on Gemini South paving the way. GeMS has been routinely producing corrected science images over an $\SI{85}{\arcsecond}\times\SI{85}{\arcsecond}$ field since 2013. Solar MCAO has an even longer history,\cite{Berkefeld2005,Berkefeld2006} although early experiments were hampered by problems with loop stability. The 3 DM MCAO system Clear at the Goode Solar Telescope has provided high-order AO correction over a \SI{30}{\arcsecond}$times$\SI{30}{\arcsecond} at visible wavelengths since 2016.\cite{Schmidt2017}

The combination of one or more DMs at a very high altitude and the small subaperture sizes in solar MCAO can lead to an unstable loop in the presence of moderate to strong seeing. Table \ref{tab:mcaoparams} records the salient differences between night-time and solar MCAO that affect the loop stability.
\begin{table}[H]
  \centering
  \caption{Comparison of relevant parameters for the MCAO systems on Gemini South and Gregor.\cite{Gregor}}
  \begin{tabular}{l|rr}\toprule[1pt]
    Parameter &  Night-time & Solar \\ \midrule
    Highest DM conjugate altitude & \SI{9}{\kilo\meter} & \SI{25}{\kilo\meter} \\ 
	Subaperture size & \SI{0.50}{\meter} & \SI{0.10}{\meter} \\ \bottomrule[1pt]
  \end{tabular}
  \label{tab:mcaoparams}
\end{table}
In addition, solar telescopes typically point at lower elevations than night-time telescopes, which have the luxury of observing targets when they are higher in the sky. If the wavefront sensors (WFSs) are conjugate to the ground ({\it i.e.}, in the pupil plane), then the propagation between the high altitude DM and the ground leads to a pupil distortion. A distorted pupil leads to a change in registration between the WFSs and the DM. If the interactuator spacing is matched to the size of the subaperture, a pupil distortion of half a subaperture leads to an unstable loop, because an actuator poke will be incorrectly reconstructed as originating from a neighboring actuator. 

\begin{figure}[H]
  \centering
  \includegraphics[height=6cm]{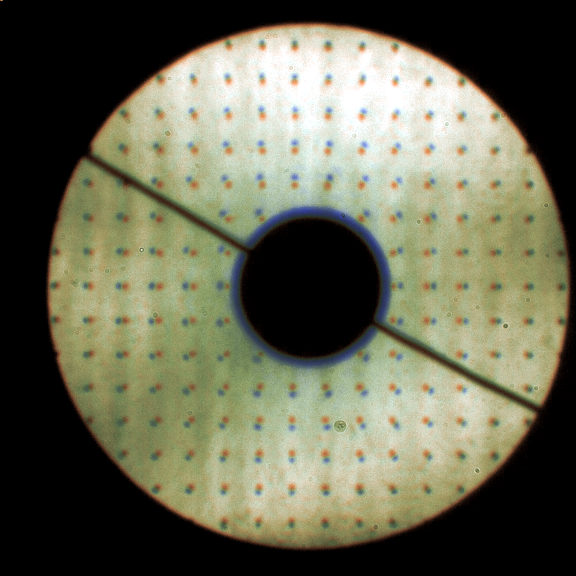} 
  \includegraphics[height=6cm]{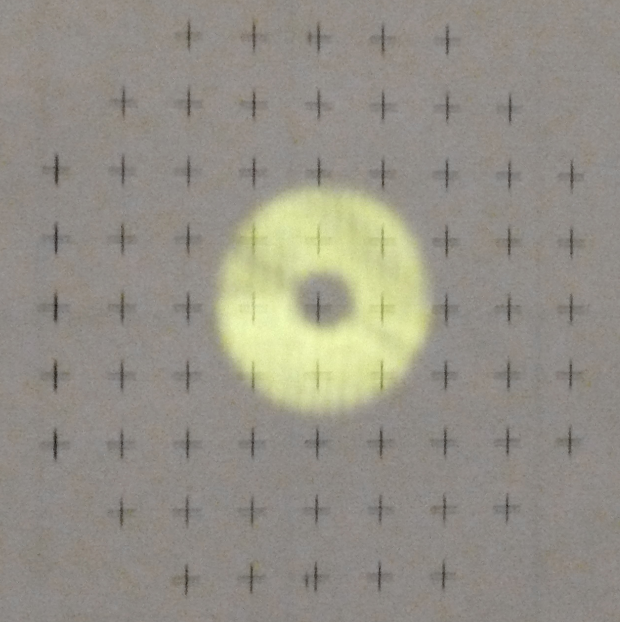} 
  \caption{Photographic illustration of pupil distortion: the same grid pattern is shown with the DMs flat and with a \SI{400}{\nano\meter} poke of the central actuator on high-altitude DM (left). The region corresponds to an area spanning approximately $3\times3$ actuators marked with a cross (right).\cite{Schmidt2010}}
  \label{fig:gregorbench}
\end{figure}

The distorting effect of the high-altitude DMs on the exit pupil in MCAO was first noticed by von der Lühe while observing rapid intensity fluctuations in the MCAO corrected image plane at the VTT.\cite{vonderLuhe2004} Later, it was realized that the ever changing shape of the high-altitude DM that continuously changes the registration between the pupil DM and the WFS could potentially pose a challenge to MCAO control. In a 2010 paper, Schmidt {\em et al.} demonstrated the dynamic misregistration using ray tracing and experimentally. Figure \ref{fig:gregorbench} illustrates the pupil distortion and the associated dynamic misregistration of the the Gregor MCAO system by poking the central actuator of the high altitude DM.\cite{Schmidt2010}  
In the second generation of solar MCAO systems at Gregor and Clear at the Goode Solar Telescope, it was a goal to mitigate the misregistration issue. In the Gregor system, a high-order on-axis WFS was placed between the pupil DM and the high-altitude DMs, and a low-order multi-directional WFS was placed behind all DMs. The same scheme was available in Clear as one of six different WFS and DM concepts implemented. This control strategy is stable and produced AO corrected images.\cite{Schmidt2014,Schmidt2016,Schmidt2017} 

In Section \ref{sec:simulationcodes}, we compare existing AO simulation and describe how they propagate the wavefront. To our knowledge, there is no existing end-to-end Monte-Carlo simulation tool that models solar MCAO and reproduces this effect. We considered three different ways to model the pupil distortion:
\begin{itemize}
\item Fresnel propagation
\item Ray tracing (using the start point and the gradient of the ray to determine the end point)
\item The irradiance transport equation along with the associated wavefront transport equation\cite{Teaguegreen}
\end{itemize}
The quickest way to create a new simulation tool with the required properties was to adapt PROPER,\cite{Krist2007} an optical propagation code based on Fresnel propagation (Section \ref{sec:propao}). 

In this paper, we develop PropAO, the first AO simulation tool (to our knowledge) to model the effect of pupil distortion. PropAO takes advantage of the optical propagation library PROPER and uses Fresnel propagation to propagate the amplitude and phase of an incoming wave through the atmosphere and the MCAO system. The resulting wavefront is analyzed by the WFSs and also used to evaluate the corrected image quality. We are able to reproduce the problem of pupil distortion and test novel non-linear reconstruction strategies that take the distortion into account and produce well-corrected wide field images.

Existing AO systems use linear wavefront reconstruction techniques, where the estimates wavefront is a linear combination of the measured wavefront derivatives. In Section \ref{sec:reconstructors}, we propose new reconstructors to combat the pupil distortion problem. Simulation results using a toy MCAO problem are presented in Section \ref{sec:simulations} and conclusions are drawn in Section \ref{sec:conclusion}.

The motivation for this work is the design of the \SI{4.2}{\meter} diameter European Solar Telescope.\cite{Montoya2015,Berkefeld2018} The current design has a very ambitious \SI{40}{\arcsecond}$\times$\SI{40}{\arcsecond} AO-corrected science field of view. In order to deliver near diffraction-limited performance at visible wavelengths over such a wide field of view, an MCAO system with three to five DMs is under consideration. The ability to model the pupil distortion is essential to the success of the EST project. The PropAO tool and the results of this work are also applicable to other solar and night-time astronomy MCAO systems.

\section{Existing simulation codes}
\label{sec:simulationcodes}
In this section, we describe publicly available and well-documented end-to-end simulation codes. We evaluate them on their suitability to model the pupil distortion of a solar MCAO system and find that none of these codes can be used to model the pupil distortion. 

\subsection{DASP}
The Durham Adaptive optics Simulation Platform (DASP) is an open-source configurable AO simulation environment written in Python.\cite{Basden,DASP} DASP offers support for tomographic reconstructors and pseudo-open loop control, so it is well suited for MCAO simulations. The turbulence propagation is performed either by simple integration of the wavefront along the line of sight or using Fresnel propagation. Only the former option is currently available for propagation through the DM to the WFSs. Implementing this using Fresnel propagation, while not a lot of work, is not in the pipeline.\cite{BasdenPersonalComms} 

\subsection{YAO}
A full-featured end-to-end Monte-Carlo simulation tool, known as yorick adaptive optics (YAO) and written by Francois Rigaut, is used extensively in the AO community.\cite{RigautYAO,YAO} The wavefront propagation is performed by summing the wavefront through each atmospheric layer or DM surface in the direction of each WFS or science target. Fresnel propagation is not implemented. The pupil distortions are not modeled.

\subsection{OOMAO}
OOMAO is a Matlab based AO simulation tool that is highly configurable. The wavefront propagation is performed in the same way as YAO, so the pupil distortion is not modeled, either.

\subsection{CAOS}
The Software Package CAOS is a subset of the CAOS Problem-Solving Environment and consists of an IDL based AO simulations tool.\cite{Carbillet2017,CAOS} It uses linear wavefront propagation to calculate the wavefronts. It was claimed in 2005 that ``a Fresnel module for atmospheric turbulence has been written and will be part of a forthcoming release.''\cite{Carbillet2005} This claim is still true today! Work on Fresnel propagation through turbulence was paused, has recently restarted but is not yet ready for release.\cite{CarbilletPersonalComms} It is possible to also implement Fresnel propagation between DMs at different conjugate altitudes, but this functionality is not planned. 

\subsection{CEO}
CEO (an acronym for Cuda Engined Optics) is a simulation tool written for the Giant Magellan Telescope.\cite{Conan2015,CEO} This open-source software applies ray tracing in CUDA and has a Python interface. It has the capability to apply ray tracing through the atmosphere and the telescope, but not through an AO system. This capability would be needed to model the pupil distortions.

\begin{figure}[H]
  \centering
  \includegraphics[width=5.6cm,height=4.2cm]{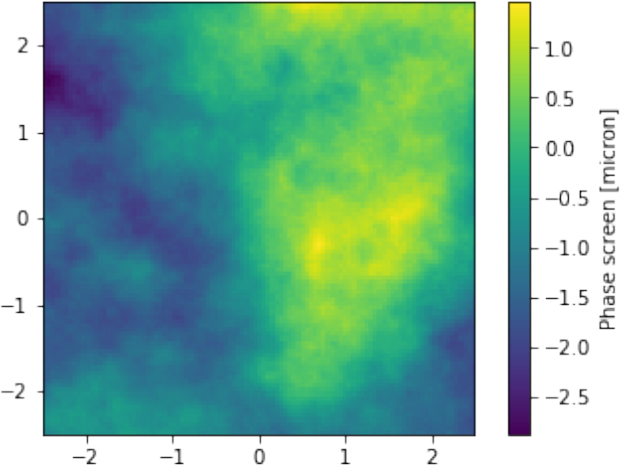} 
  \includegraphics[width=5.84cm,height=4.2cm]{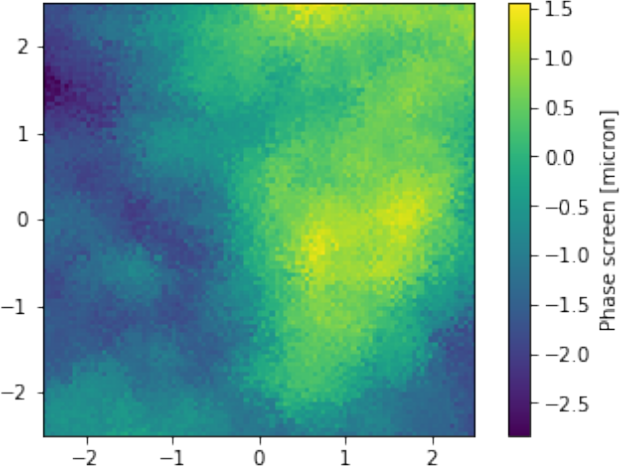} 
  \includegraphics[width=5.219cm,height=4.2cm]{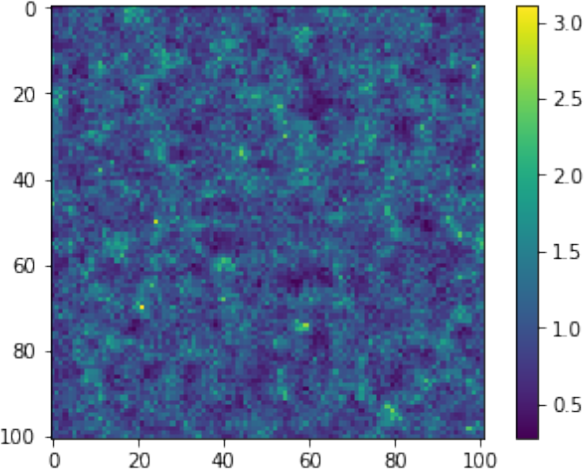}
  \caption{Phase before propagation (left) and phase (center) and amplitude (right) after propagation.}
  \label{fig:ceopropagation}
\end{figure}

Figure \ref{fig:ceopropagation} shows how the wavefront propagation leads to variations in wavefront and amplitude using ray tracing in CEO.

\section{PropAO: an extension to the PROPER optical propagation tool}
\label{sec:propao}
\begin{figure}[htb]
  \centering
  \includegraphics[width=6.0cm]{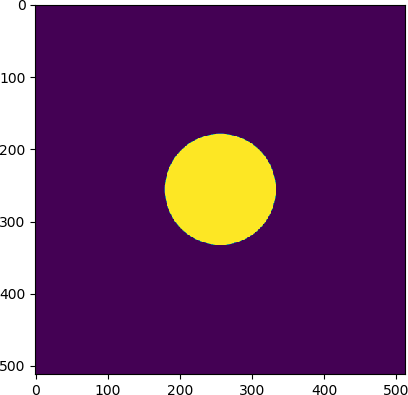} 
  \includegraphics[width=6.0cm]{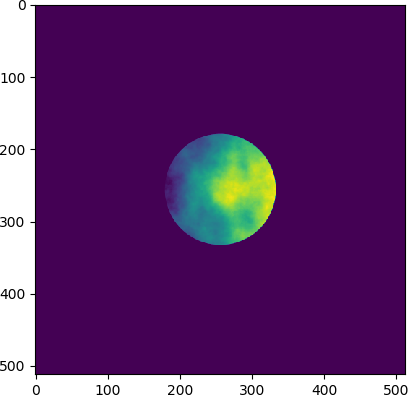}
  \includegraphics[width=6.0cm]{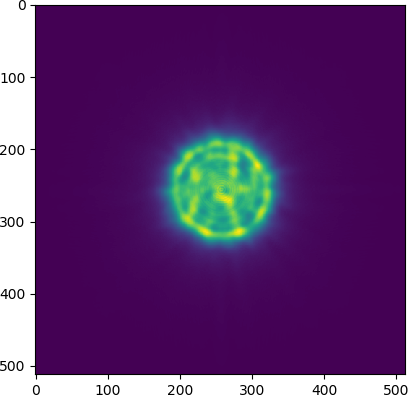} 
  \includegraphics[width=6.0cm]{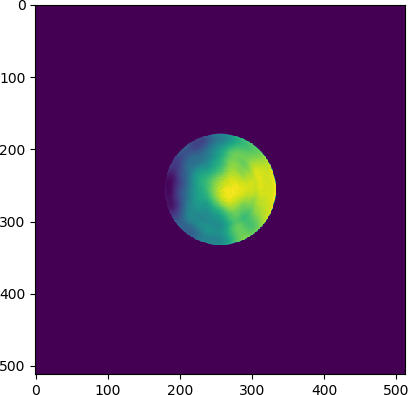}
  \caption{Initial amplitude and phase (top) and final amplitude and phase (bottom) following a \SI{20}{\kilo\meter} propagation using the PROPER code.}
  \label{fig:properpropagation}
\end{figure}

The PROPER optical propagation code, available in IDL, Matlab and Python, is a library of routines for optical propagation.\cite{Krist2007,PROPER} PROPER uses full Fresnel propagation at every step and is suitable for modeling the complex amplitude of a wave propagating through the atmosphere and an MCAO system. There is support for pupil stops, DMs, and point-spread function (PSF) generation. An example of Fresnel propagation using PROPER is shown in Figure \ref{fig:properpropagation}, where a Kolmogorov phase screen is propagated \SI{20}{\kilo\meter} in broadband
light over a the range of \SI{400}{\nano\meter} to \SI{800}{\nano\meter} nm. The amplitude and the phase evolve as the wave propagates.

In this section, we show how we extend the functionality of PROPER into an end-to-end simulation tool. PropAO uses the Python 3 version of the PROPER optical propagation tool as a starting point and adds new features missing in PROPER, such as atmospheric turbulence and WFSs, in order to create an end-to-end MCAO simulation tool. As will become evident, the development process consisted a tool that mimicked the behavior of YAO as closely as possible in order to quickly and efficiently debug and verify the new code using a simple single conjugate AO system. Once we were confident in the results, the simulation was extended to an MCAO with multiple DMs and WFSs.

\subsection{Atmospheric turbulence}
PROPER has the capability of inserting amplitude or phase screens anywhere in the optical train, but no function to create Kolmogorov phase screens. Atmospheric phase screens were generated using YAO to avoid writing code unnecessarily, and to make it easier to compare the results between the two codes. The optical propagation begins at the top of the atmosphere, where the first atmospheric phase screen is inserted. The complex amplitude is propagated to the next atmospheric phase screen, where the phase corresponding to the second atmospheric phase screen is added. This procedure is repeated until we reach the top of the telescope.

\subsection{Telescope}
The telescope is modeled as an ideal telescope. The complex amplitude of the incoming wave is multiplied by a circular function representing the exit pupil of the telescope by calling a PROPER function. In the future, we could include a realistic optical model of the telescope.

\subsection{Deformable mirrors}
The first DM is conjugate to the pupil, so no propagation from the telescope exit pupil is required. The complex amplitude is multiplied by the phase induced by the DM and the wave is propagated to the first altitude DM. This process was repeated for all DMs. Finally, the wave is progated back to the pupil plane, where the wavefront sensors (WFSs) and the science instruments are situated.

The functions in PROPER used to mode DMs were initially used. However, two limitations drove us to write our own DM module: flexibility and speed. The DM influence functions in PROPER are hard coded to a particular function that is loaded as a FITS file. We wanted to be able to specify the same influence functions as YAO initially, and then to be able to use the influence functions corresponding to a real DM. The DM implementation in PROPER was also prohibitively slow. 

The new implementation for the DMs is as follows. First, the user can define an actuator influence function or use one of the pre-programmed ones. The options include the sinc interpolator, the bilinear intepolator, an approximation to the measured influence function of a Xinetics actuator, and the same function that is programmed in YAO. The two-dimensional actuator commands are convolved with the influence function. The convolution takes place as a multiplication in the Fourier domain using FFTs.

\subsection{Wavefront sensors}
There are no WFSs in PROPER. Shack-Hartmann WFSs were implemented in the following manner. The complex amplitude at the pupil plane is subdivided by the lenslet array and each subaperture is propagated onto the pupil plane to form an image. The centroid of the image on each subaperture is computed and recorded. 

\subsection{Wavefront reconstruction and control}
The main purpose of the tool is to investigate wavefront reconstruction and control strategies. The tool supports closed-loop and open-loop (where the WFSs do not see the DMs) modes as well as pseudo open-loop control, as described in Section \ref{sec:reconstructors}. 

\subsection{Science instrument}
The science instrument is modeled as an imager which takes an image of a point source and uses this image to estimate the Strehl ratio at the science wavelength(s). There is the option of including a pupil stop in the science camera.

\subsection{Use of GPUs with the CuPY library} 
\label{sec:gpu}
Following the completion of the first operational version PropAO, it was evident from modeling a \SI{1}{\meter} diameter telescope that the simulations could not be scaled to a \SI{4.2}{\meter} diameter telescope like the European Solar Telescope due to the computational load. Since most of the computations consist of FFTs and matrix multiplications, PropAO is an ideal candidate for the use of Graphical Processing Units (GPUs). The CuPY library is a Python library that uses CUDA acceleration for the heavy numerical computations.\cite{CuPY} It is almost a like-to-like replacement for the NumPY library, which is used in PROPER. By making a small number of changes (fewer than ten different lines of code), we were able to convert the computations in PropAO from CPU to GPU. The simulation time using an off-the-shelf laptop (Dell Precision 7710 with an NVIDIA Quadro M4000M GPU on a laptop) decreased by a factor of 100! As an example, a 3 DM system on a \SI{4}{\meter} telescope with 5 $32\times32$ subaperture WFSs runs at 2.5 iterations/second. 

\section{Wavefront reconstruction and control}
\label{sec:reconstructors}

In this section, we describe a number of different approaches to the problem of wavefront reconstruction and control, which are subsequently evaluated using PropAO simulations.

\subsection{Closed-loop control with regularized least-squares reconstructor}
\begin{figure}[H]
  \centering
  \includegraphics[height=3.3cm]{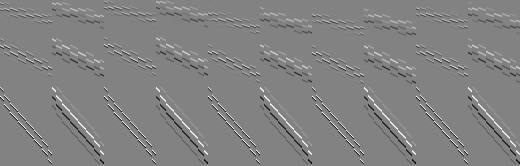} 
  \includegraphics[height=3.3cm]{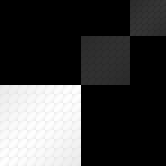}
  \includegraphics[height=3.3cm]{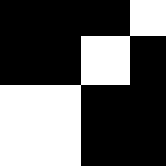} 
  \caption{Interaction matrix (left), covariance matrix (center) and piston penalization matrix (rigth) for a 3 DM, 5 WFS system.}
  \label{fig:imat}
\end{figure}

The standard closed-loop reconstructor is what is used on almost all AO systems. An interaction matrix is created by poking the actuators one by one. The reconstructor is computed as a regularized inverse of the interaction matrix, either using a singular value decomposition, a regularized least-squares inversion or a modal basis set. 
We use the regularized least-squares reconstructor, $R$:
\begin{equation}
\label{eq:lsrec}
R = (H^T C_{ nn }^{ -1 }H + \alpha C_\phi^{ -1 }+P)^{ -1 }H^T C_{ nn }^{ -1 }
\end{equation}  
where $H$ is the interaction matrix, $P$ is a matrix that penalizes piston for each DM, $C_{nn}$ is the noise covariance matrix, $C_\phi$ is the covariance matrix of the uncorrected turbulence and $\alpha$ is a regularization constant. Figure \ref{fig:imat} plots the interaction matrix, covariance matrix and piston penalization matrix for a 3 DM, 5 WFS MCAO system. The noise covariance matrix is simply a diagonal matrix.  
The wavefront residual, $u$, is given by the matrix multiplication of the reconstructor, $R$, and the centroids, $s$:
\begin{equation}
u=Rs
\end{equation}
A leaky integrator with leak $l$=0.01 loop gain $k$=0.6 is used to update the actuator commands, $a$ at time $n$:
\begin{equation}
a[n]=(1-l)a[n-1]+ku[n].
\end{equation}
This reconstruction and control approach can lead to instabilities because the high altitude DMs will cause a shift in the registration between the actuators of the lower altitude DM and the WFS lenslets.

\subsection{Closed-loop control with iterative interaction matrix and regularized least-squares reconstructor}
The interaction matrix is calculated by poking actuators one by one away from their nominal positions and measuring the centroids. However, the effect of each actuator on the WFS depends on the actuator commands of the mirrors that follow it. For example, the effect that the ground-layer DM has on the WFS centroids depends on the shape of the mid- and high-altitude DMs. In this control strategy, we calculate the interaction matrix iteratively by poking actuators referenced to the current DM commands. Each time the interaction matrix is calculate (once every ten iterations in the simulations presented in this paper), the reconstructor is recomputed.

\subsection{Pseudo open-loop control with regularized least-squares reconstructor}
\label{sec:polc}
The reconstructor in Eq. (\ref{eq:lsrec}) can be used in pseudo open-loop control. Pseudo-open loop control uses the open-loop centroids, $s -Ha$ to calculate the residuals:
\begin{eqnarray}
u &=& R(s + Ha) - a \nonumber \\ 
u &=& Rs +({RH}-I)a
\end{eqnarray}
where $I$ is the identity matrix.

\subsection{Non-linear pseudo open-loop control with regularized least-squares reconstructor}
The pseudo open-loop control described in Section \ref{sec:polc} relies on a linear relationship between the actuator commands and the centroids to compute the open-loop centroids. For strong turbulence, this relationship does not hold due to the pupil distortion effect. A more accurate way to compute the open-loop centroids is to propagate a wave from the top of the telescope through to the WFSs and measure the centroids. This propagation takes place once per iteration. 
 
\subsection{Separate control of ground-layer DM and high-altitude DMs}
In some solar MCAO systems, a high-order on-axis WFS is used to control the ground-layer DM, while lower-order off-axis WFSs are used to drive the mid- and high-altitude DMs.\cite{Schmidt2016} The on-axis WFS does not see the other DMs, so the registration between the WFS and the ground-layer DM is not affected by the other DMs. Here, the optical setup is different, not just the control law!

\section{Simulations}
\label{sec:simulations}
Simulations were run using new simulation tool and compared with the same
simulations run in YAO in order to understand the effects of Fresnel propagation through the atmosphere and AO system. 

\subsection{Simulation parameters}
The simulations were run using a \SI{1}{\meter} diameter telescope with no central
obscuration. In all cases, there are $8\times8$ subapertures across the pupil. Two different atmospheric profiles are used, one with 70\% of the turbulence near the ground (Profile 1) and another one with 90\% (Profile 2). The turbulence parameters are tabulated in Table \ref{tab:atmosphere}. 

\begin{table}[H]
  \centering
  \caption{Turbulence profiles used in the simulations.}
  \begin{tabular}{l|l|rrrrrr}
  \toprule[1pt]
  &Elevatiion (\SI{}{\meter}) & 0 & 1000 & 2000 & 4000 & 8000 & 16000  \\
  Profile 1 & Turbulence fraction & 0.55 & 0.15 & 0.05 & 0.06 & 0.07 & 0.12 \\
  Profile 2 & Turbulence fraction & 0.63 & 0.25 & 0.02 & 0.02 & 0.03& 0.05 \\ 
  &Wind speed (\SI{}{\meter\per\second}) & 5.6 & 6.25 & 7.57 & 13.31 & 19.06 & 12.14 \\
  &Wind direction (\SI{}{\degree}) & 190 & 270 & 350 & 17 & 29 & 66   \\
  \bottomrule[1pt]
  \end{tabular}
  \label{tab:atmosphere}
\end{table}
The outer scale of turbulence used in the simulations is a few meters, which is not
realistic. This occurs because the phase screens are periodic in order to avoid issues with discontinuities when propagating with the Fast Fourier Transform (FFT). The effect of the outer scale on AO-corrected performance is expected to be small, but needs to be considered. 

\subsection{Seeing-limited simulations}
Simulations were run with no correction as a sanity check, and compared with a YAO simulation using the same parameters. There were 3000 iterations with a \SI{10}{\milli\second} exposure time for a total integration time of \SI{30}{\second}. There is excellent consistency between the YAO and the PropAO results (Table \ref{tab:uncorrected}), which is not suprising since the same turbulence phase screens are used. The on-axis Strehl ratio at \SI{500}{\nano\meter} is used as the science metric.  
\begin{table}[H]
  \centering
  \caption{Strehl ratio at \SI{500}{\nano\meter} from YAO and PropAO simulations of uncorrected turbulence.}
  \begin{tabular}{c|c|rrr|rrr}
  \toprule[1pt]
    & & \multicolumn{3}{c|}{Profile 1} & \multicolumn{3}{c}{Profile 2} \\
   \midrule
    & Elev. & $r_0 = $ \SI{8}{\centi\meter} & $r_0 = $ \SI{15}{\centi\meter} & $r_0 = $ \SI{25}{\centi\meter} & $r_0=$ \SI{8}{\centi\meter}& $r_0=$ \SI{15}{\centi\meter} & $r_0=$ \SI{25}{\centi\meter} \\
   \midrule
   YAO & \SI{20}{\degree} & 0.0020 & 0.0061 & 0.0160 & 0.0020 & 0.0062 & 0.0164 \\
   \rowcolor{Gray} PropAO  & \SI{20}{\degree} & 0.0018 & 0.0063 & 0.0179 & 0.0018 & 0.0063 & 0.0176 \\
     \midrule
  YAO & \SI{45}{\degree} & 0.0043 & 0.0145&0.0394&0.0044&0.0146&0.0405\\
   \rowcolor{Gray}  PropAO & \SI{45}{\degree}&  0.0042 & 0.0156 & 0.0432 & 0.0043 & 0.0153 & 0.0415 \\
     \midrule
YAO & \SI{90}{\degree}  & 0.0067 & 0.0232 & 0.0607 & 0.0063
& 0.0221 & 0.0593 \\
\rowcolor{Gray} PropAO & \SI{90}{\degree} & 0.0067 & 0.0232 & 0.0607 & 0.0063 & 0.0221 & 0.0593\\
  \bottomrule[1pt]
  \end{tabular}
  \label{tab:uncorrected}
\end{table}
\subsection{Single conjugate adaptive optics simulations}
A simple single conjugate adaptive optics (SCAO) system was simulated using both codes. A ground-layer DM with $9\times9$ actuators across the pupil in the so-called Fried configuration was used. The WFS exposure time was \SI{1}{\milli\second}, and 1000 iterations were run for a total simulation time of \SI{1}{\second}. All of the simulations in this section are performed using pseudo open-loop control with an optimized value of $\alpha$ in the least-squares inversion of Eq. (\ref{eq:lsrec}). The turbulence was moved to the ground in order to avoid scintillation effects, which are not modeled in YAO. The results are tabulated in Table \ref{tab:scao}.
\begin{table}[H]
  \centering
  \caption{Strehl ratio at \SI{500}{\nano\meter} from YAO and PropAO simulations of an SCAO system with all of the turbulence near the ground.}
  \begin{tabular}{cc|rrr|rrr}
  \toprule[1pt]
    & & \multicolumn{3}{c|}{Profile 1} & \multicolumn{3}{c}{Profile 2} \\
   \midrule
       & Elev. & $r_0 = $ \SI{8}{\centi\meter} & $r_0 = $ \SI{15}{\centi\meter} & $r_0 = $ \SI{25}{\centi\meter} & $r_0=$ \SI{8}{\centi\meter}& $r_0=$ \SI{15}{\centi\meter} & $r_0=$ \SI{25}{\centi\meter} \\
   \midrule
   YAO & \SI{20}{\degree} & 0.093 & 0.443 & 0.706 & 0.098 & 0.452 & 0.712 \\
   \rowcolor{Gray} PropAO  & \SI{20}{\degree} & 0.089 & 0.421 & 0.690 & 0.102 & 0.442 & 0.703 \\
     \midrule
  YAO & \SI{45}{\degree}  & 0.321 & 0.673 & 0.844 & 0.321 & 0.674 & 0.845 \\
   \rowcolor{Gray}  PropAO & \SI{45}{\degree}& 0.305 & 0.657 & 0.832 & 0.325 & 0.672 & 0.840 \\
     \midrule
  YAO & \SI{90}{\degree}  &  0.435 & 0.745 & 0.882 & 0.437 & 0.747 & 0.882 \\
\rowcolor{Gray} PropAO & \SI{90}{\degree} & 0.431 & 0.741 & 0.876 & 0.451 & 0.754 & 0.882 \\
  \bottomrule[1pt]
  \end{tabular}
  \label{tab:scao}
\end{table}
The theoretical decrease in Strehl ratio, $S$ due to scintillation is given by the Mar\'echal approximation for sufficiently small amplitude vatiations:
\begin{equation}
S = \exp [-\sigma _{\chi }^2]
\end{equation}
where the scintillation is calculated as\cite{SasielaBook}
\begin{equation}
\sigma _{\chi }^2 = 0.5631k^{7/6}\sec^{11/6}(\zeta )\int dz C_n^2(z) z^{5/6}.
\end{equation}
The symbol $\zeta$ represents the zenith angle, and $k$ is the wave number, $2\pi /\lambda$. 

PropAO simulations were repeated with the turbulence phase screens at their nominal value, and the decrease in Strehl ratio was compared to the expected decrease in Strehl due to scintillation (Table \ref{tab:scintillation}). In general, both sets of results show the same trends, but the fit is not perfect. Part of the reason is that the scintillation also has an effect on the WFS, so the Strehl ratio is worse than predicted when the scintillation is strong. The results at an elevation of \SI{20}{\degree} are omitted because the Mar\'echal approximation does not hold.
\begin{table}[H]
  \centering
  \caption{Strehl ratio at \SI{500}{\nano\meter} from YAO and PropAO simulations of an SCAO system with all of the turbulence near the ground.}
  \begin{tabular}{cc|rrr|rrr}
  \toprule[1pt]
      & & \multicolumn{3}{c|}{Profile 1} & \multicolumn{3}{c}{Profile 2} \\
   \midrule
       & Elev. & $r_0 = $ \SI{8}{\centi\meter} & $r_0 = $ \SI{15}{\centi\meter} & $r_0 = $ \SI{25}{\centi\meter} & $r_0=$ \SI{8}{\centi\meter}& $r_0=$ \SI{15}{\centi\meter} & $r_0=$ \SI{25}{\centi\meter} \\
     \midrule
  Analytic & \SI{45}{\degree} & 0.891 & 0.960 & 0.983 & 0.942 & 0.979 & 0.991 \\
\rowcolor{Gray} PropAO & \SI{45}{\degree} & 0.790  & 0.951 & 0.994 & 0.898 & 0.987 & 0.998 \\
     \midrule
  Analytic & \SI{90}{\degree} & 0.941 & 0.978 & 0.991 & 0.969 & 0.989 & 0.995 \\
  \rowcolor{Gray} PropAO & \SI{90}{\degree} & 0.929 & 0.991 & 0.998 & 0.970 & 0.996 & 0.999 \\
  \bottomrule[1pt]
  \end{tabular}
  \label{tab:scintillation}
\end{table}
\subsection{Multiconjugate adaptive optics simulations}
\label{sec:mcao}
Simulations were run for a toy 3-DM MCAO system using the same \SI{1}{\meter} telescope. The DM parameters are as described in Table \ref{tab:dmparams}.
\begin{table}[H]
  \centering
  \caption{Description of the DMs used in the MCAO simulations.}
  \begin{tabular}{r|r|r|r}
  \toprule[1pt]
  DM number & Altitude & Interactuator spacing &Actuators across pupil \\ \midrule
  0 & \SI{0}{\meter} & \SI{125}{\milli\meter} & $9\times9$ \\
  1 & \SI{5000}{\meter} & \SI{250}{\milli\meter} & $7\times7$ \\
  2 & \SI{12000}{\meter} & \SI{375}{\milli\meter} & $6\times6$ \\
	\bottomrule[1pt]
  \end{tabular}
  \label{tab:dmparams}
\end{table}
There are five identical WFSs with $8\times8$ subapertures. The guide stars are modeled as point sources located at $[\SI{0}{\arcsecond},\SI{0}{\arcsecond}]$, $[\pm \SI{5}{\arcsecond},\pm \SI{5}{\arcsecond}]$. The science field of view is a $\SI{10}{\arcsecond}\times\SI{10}{\arcsecond}$ square. The results from the YAO simulations using pseudo open-loop control are displayed in Table \ref{tab:yaomcaosims}. All of these YAO simulations exhibited loop stability and improved the image quality; this is not surprising, because YAO does not model the pupil distortion. It is clear that some configurations, such as observing at an elevation of \SI{20}{\degree} in anything but the best seeing, are not really suitable for wide-field MCAO correction. 
\begin{table}[H]
  \centering
  \caption{Strehl ratio at 500 nm from MCAO simulations using YAO with psedo open-loop control.}
  \begin{tabular}{c|rrr|rrr}
  \toprule[1pt]
     & \multicolumn{3}{c|}{Profile 1} & \multicolumn{3}{c}{Profile 2} \\
   \midrule
    Elev. & $r_0 = $ \SI{8}{\centi\meter} & $r_0 = $ \SI{15}{\centi\meter} & $r_0 = $ \SI{25}{\centi\meter} & $r_0=$ \SI{8}{\centi\meter}& $r_0=$ \SI{15}{\centi\meter} & $r_0=$ \SI{25}{\centi\meter} \\
     \midrule
\SI{20}{\degree} & 0.005±0.001 & 0.037±0.016 & 0.152±0.058 & 0.008±0.002 & 0.092±0.032 & 0.316±0.065 \\
     \midrule
\SI{45}{\degree} & 0.088±0.023 & 0.417±0.040 & 0.687±0.028 & 0.189±0.021 & 0.548±0.022 & 0.773±0.013 \\
     \midrule
\SI{90}{\degree} & 0.268±0.017 & 0.629±0.015 & 0.820±0.008 & 0.361±0.011 & 0.699±0.008 & 0.858±0.004 \\
  \bottomrule[1pt]
  \end{tabular}
  \label{tab:yaomcaosims}
\end{table}
These simulations were repeated in PropAO using a variety of reconstruction strategies, described in Section \ref{sec:reconstructors}: 
\begin{itemize}
\item Regularized least-squares reconstructor operating in closed-loop (CL)
\item Regularized least-squares reconstructor with an iterative interaction matrix operating in closed-loop (Iter CL)
\item Regularized least-squares reconstructor operating in pseudo open-loop (POLC)
\item Regularized least-squares reconstructor with non-linear pseudo open-loop (NL POLC)
\end{itemize} 

The performance achieved was very sensitive to the level of regularization (the value of the $\alpha$ parameter) used in the reconstructor, with the regularization adjusted for each simulation in order to maximize the Strehl ratio. For strong turbulence cases, a high level of regularization was needed to maintain loop stability.  
\begin{table}[H]
  \centering
  \caption{Strehl ratio at \SI{500}{\nano\meter} from MCAO simulations using PropAO with varying reconstruction and control strategies.}
  \begin{tabular}{cc|rrr|rrr}
  \toprule[1pt]
    & & \multicolumn{3}{c|}{Profile 1} & \multicolumn{3}{c}{Profile 2} \\
   \midrule
 & Elev. & $r_0 = $ \SI{8}{\centi\meter} & $r_0 = $ \SI{15}{\centi\meter} & $r_0 = $ \SI{25}{\centi\meter} & $r_0=$ \SI{8}{\centi\meter}& $r_0=$ \SI{15}{\centi\meter} & $r_0=$ \SI{25}{\centi\meter} \\
   \midrule
   CL & \SI{20}{\degree}  & --- & 0.045±0.012 & 0.206±0.050 & --- & 0.119±0.034 & 0.385±0.047 \\
   Iter CL & \SI{20}{\degree} & 0.005±0.001 & 0.046±0.021 & 0.206±0.048 & 0.008±0.002 & 0.119±0.031 & 0.375±0.041 \\
   POLC & \SI{20}{\degree} & --- &0.048±0.019 & 0.206±0.043 & --- & 0.128±0.029 & 0.384±0.038 \\
   NL POLC & \SI{20}{\degree} & ---& 0.046±0.016 & 0.200±0.042 & --- & 0.116±0.025 & 0.367±0.038 \\
     \midrule
   CL & \SI{45}{\degree}  &0.064±0.012&0.402±0.035&0.663±0.024&0.142±9.014&0.536±0.021&0.754±0.013\\
   Iter CL & \SI{45}{\degree} & 0.068±0.015 & 0.388±0.035 & 0.658±0.025 & 0.142±0.010 & 0.520±0.022 & 0.729±0.015 \\
   POLC & \SI{45}{\degree} & 0.075±0.015 & 0.419±0.035 & 0.675±0.024 & 0.182±0.020 & 0.550±0.021 & 0.763±0.021 \\
   NL POLC & \SI{45}{\degree} & 0.075±0.017 & 0.397±0.034 & 0.660±0.024 & 0.152±0.017 & 0.523±0.021 & 0.747±0.013 \\
   \midrule
   CL & \SI{90}{\degree}  &0.226±0.014&0.587±0.012&0.783±0.007&0.267±0.012&0.670±0.009&0.831±0.006 \\
   Iter CL & \SI{90}{\degree} & 0.194±0.012 & 0.575±0.012 & 0.784±0.007 & 0.288±0.010 & 0.662±0.008 &0.816±0.008 \\
   POLC & \SI{90}{\degree} & 0.249±0.015 & 0.618±0.013 & 0.806±0.007 & 0.358±0.011 & 0.696±0.008 &0.849±0.005 \\
   NL POLC & \SI{90}{\degree} & 0.224±0.014 & 0.600±0.013 & 0.797±0.008 & 0.318±0.011 & 0.673±0.008 & 0.838±0.005 \\
  \bottomrule[1pt]
  \end{tabular}
  \label{tab:mcaoclpolc}
\end{table}
The results in Table \ref{tab:mcaoclpolc} shows that all of the methods could close the loop successfully, except in the worst case where the turbulence was strong and the elevation was low. The key to loop stability was a high level of regularization, but this limits the attainable atmospheric correction. 

The best control strategy is to use linear pseudo open-loop control. The non-linear control strategies did not improve the performance across the board and often made things worse. 

\subsection{Multiconjugate adaptive optics simulations with decoupled DM control}
The simulation is run using the same DM configuration is used, but two sets of WFSs are used:
\begin{itemize}
\item a pick-off after the ground-layer DM that feeds an on-axis WFS with 8x8 subapertures across the pupil. 
\item A second pick-off following the three DMs feeds the four off-axis WFSs, which have 4x4 subapertures across the pupil.   
\end{itemize}
The correction of the ground-layer DM is decoupled from the high-altitude DMs, as described in Schmidt {\em et al}.\cite{Schmidt2017} 

Simulations run using YAO and PropAO show consistent results, with the PropAO values lower than the YAO results mostly due to scintillation (Table \ref{tab:decoupled}). The Strehl ratios are much lower than those obtained in Section \ref{sec:mcao}. The decoupled DM control strategy has lower performance because the on-axis wavefront is not the same as the wavefront at the ground. This strategy should not be used. 
\begin{table}[H]
  \centering
  \caption{Strehl ratio at \SI{500}{\nano\meter} from YAO and PropAO simulations using MCAO with decoupled control.}
  \begin{tabular}{cc|rrr|rrr}
  \toprule[1pt]
    & & \multicolumn{3}{c|}{Profile 1} & \multicolumn{3}{c}{Profile 2} \\
   \midrule
    & Elev. & $r_0 = $ \SI{8}{\centi\meter} & $r_0 = $ \SI{15}{\centi\meter} & $r_0 = $ \SI{25}{\centi\meter} & $r_0=$ \SI{8}{\centi\meter}& $r_0=$ \SI{15}{\centi\meter} & $r_0=$ \SI{25}{\centi\meter} \\
   \midrule
   YAO &\SI{20}{\degree} & 0.004±0.001 & 0.025±0.006 & 0.101±0.022 & 0.007±0.001 & 0.057±0.012 & 0.230±0.034 \\
   \rowcolor{Gray}PropAO & \SI{20}{\degree}  & 0.001±0.000 & 0.018±0.004 & 0.086±0.033 & 0.001±0.000 & 0.038±0.011 & 0.187±0.045 \\
     \midrule
YAO &\SI{45}{\degree} & 0.033±0.019 & 0.260±0.049 & 0.555±0.042 & 0.101±0.022 & 0.436±0.030 & 0.700±0.020 \\
   \rowcolor{Gray}PropAO  & \SI{45}{\degree} & 0.019±0.005 & 0.210±0.032 & 0.516±0.033 & 0.063±0.011 & 0.379±0.025 & 0.660±0.019 \\
     \midrule
YAO & \SI{90}{\degree} & 0.097±0.028 & 0.421±0.040 & 0.688±0.027 & 0.217±0.026 & 0.581±0.023 & 0.793±0.013 \\
  \rowcolor{Gray}PropAO & \SI{90}{\degree} & 0.056±0.012 & 0.367±0.029 & 0.660±0.022 & 0.160±0.015 & 0.544±0.022 & 0.770±0.021 \\
  \bottomrule[1pt]
  \end{tabular}
  \label{tab:decoupled}
\end{table}

\section{Conclusion}
\label{sec:conclusion}

A major stumbling block in attaining diffraction-limited images over a wide field using MCAO on solar telescopes is the dynamic change in registration between the pupil DMs and the WFSs. While this phenomenon is well understood, a simulation tool to study this problem and test potential solutions have not been available until now.

In this paper, we present a new AO simulation tool, PropAO, based on the PROPER optical propagation tool which propagates waves from plane to plane using Fresnel propagation. The Python version of the PROPER tool has been extended by adding features needed for an end-to-end MCAO simulation and has been modified to run on GPUs using the CuPY library, considering the large number of Fresnel propagations required. The simulation tool has been extensively tested by comparing outputs from simple cases to the open-source Monte-Carlo simulation tool, YAO, with excellent agreement between the codes. 

The effect of the pupil distortion on the registration between the DMs and the WFSs is readily seen in PropAO simulations. While the ultimate solution to wavefront reconstruction in the presence of pupil distortion remains elusive, it can be partially mitigated by using using a heavily regularized least-squares reconstructor in conjunction with pseudo open-loop control. This leads to a stable reconstruction for a wide range of observing conditions, and much better performance than what is attained when decoupling the control of the pupil DM and the high altitude DM using dedicated WFSs for each one.

PropAO will be used to simulate the MCAO system for the EST and to test potential implementations of wavefront control. It can also be used for any MCAO system for solar or night-time astronomy.

\section*{Acknowledgments}
This work was supported by the EST Project Office, funded by the Canary Island Government (file SD 17/01) under a direct grant awarded to EST on ground of public interest.

We extend our gratitude to Dirk Schmidt (NSO), who has been battling the dynamic misregistration problem for many years, for his advice and help with the manuscript. The authors would also like to thank John Krist (JPL) for advice on the use of PROPER and Rodolphe Conan (GMT) for the suggestion to use the CuPY library in place of the NumPY library.

\end{document}